\documentclass[10pt]{article}
\usepackage{fancyhdr}
\usepackage{geometry}
\usepackage[parfill]{parskip} 
\usepackage{graphicx}
\usepackage{amssymb,amsmath}
\usepackage{epstopdf}
\usepackage{subfigure} 
\usepackage[usenames]{color}
\usepackage[usenames]{xcolor}   \usepackage{verbatim}
\usepackage{enumitem}
\usepackage{amsmath}
\usepackage{epstopdf}
\usepackage{epsfig}
\usepackage{setspace} 
\usepackage{cite} 
\usepackage[normalem]{ulem}

\newcommand{\mycomment}[1]{{\bf{\color{red}{{}}}}} 
\newcommand{\update}[1]{{\color{red}{{}}}} 

\usepackage[labelfont=bf,labelsep=period,justification=raggedright]{caption}

\makeatletter
\renewcommand{\@biblabel}[1]{\quad#1.}
\makeatother

\begin{document}

\title{Fast and accurate imputation of summary statistics enhances evidence of functional enrichment}

\maketitle
Bogdan Pasaniuc$^{1,2}$,
Noah Zaitlen$^{3}$,
Huwenbo Shi $^2$, 
Gaurav Bhatia$^{4,5,6}$,
Alexander Gusev$^{4,5,6}$,
Joseph Pickrell$^{6,7}$,
Joel Hirschhorn$^6$,
David P Strachan$^8$,
Nick Patterson$^6$,
Alkes L. Price$^{4,5,6}$
\vspace{1cm}
\begin{enumerate}
\item Department of Pathology and Laboratory Medicine, Geffen School of Medicine, University of California Los Angeles
\item Bioinformatics Interdepartmental Program, University of California Los Angeles
\item Department of Medicine, Lung Biology Center, University of California San Francisco
\item Program in Molecular and Genetic Epidemiology, Harvard School of Public Health, 
\item Departments of Epidemiology and Biostatistics, Harvard School of Public Health, 
\item Broad Institute of Harvard and MIT, Cambridge, MA, USA.
\item Harvard Medical School, MA, USA.
\item Division of Population Health Sciences and Education, St George's,
University of London, UK.
\end{enumerate}
Correspondence: bpasaniuc@mednet.ucla.edu and aprice@hsph.harvard.edu

\doublespacing
\newpage
\section*{Abstract}
Imputation using external reference panels (e.g. 1000 Genomes) is a widely used approach for increasing power in GWAS and meta-analysis. Existing HMM-based imputation approaches require individual-level genotypes. Here, we develop a new method for Gaussian imputation from summary association statistics, a type of data that is becoming widely available.    In simulations using 1000 Genomes (1000G) data, this method recovers 84\% (54\%) of the effective sample size for common ($>$5\%) and low-frequency (1-5\%) variants  (increasing to 87\% (60\%) when summary LD information is available from target samples) versus the gold standard of 89\% (67\%) for HMM-based imputation, which cannot be applied to summary statistics.  Our approach accounts for the limited sample size of the reference panel, a crucial step to eliminate false-positive associations, and is computationally very fast.  As an empirical demonstration, we apply our method to 7 case-control phenotypes from the WTCCC data and a study of height in the British 1958 birth cohort (1958BC). Gaussian imputation from summary statistics recovers 95\% (105\%) of the effective sample size (as quantified by the ratio of $\chi^2$ association statistics) compared to HMM-based imputation from individual-level genotypes at the 227 (176) published SNPs in the WTCCC (1958BC height) data.  In addition, for publicly available summary statistics from large meta-analyses of 4 lipid traits, we publicly release imputed summary statistics at 1000G SNPs, which could not have been obtained using previously published methods, and demonstrate their accuracy by masking subsets of the data.  {We show that 1000G imputation using our approach increases  the magnitude and statistical evidence of enrichment at genic vs. non-genic loci for these traits, as compared to an analysis without 1000G imputation.  Thus, imputation of summary statistics will be a valuable tool in future functional enrichment analyses.}

\section*{Author Summary}
{Public resources of haplotypic diversity such as the 1000 Genomes are routinely leveraged to increase the number of variants tested for association in genome-wide studies through genotype imputation. Existing approaches require individual level data and are computationally demanding. We present here approaches for imputation that do not require individual level data but work directly on summary association statistics, a type of data widely available. Through extensive simulations we show that summary association imputation captures the same signal as standard individual imputation for common and to a lesser extend low-frequency variants. In addition to association studies, we show the utility of imputation of summary statistics in enrichment analyses by enhancing the signal of enrichment for functionally relevant classes of variants as compared to analyses over non-imputed data. }  

\section*{Introduction}

Genome-wide association studies (GWAS) are the prevailing approach for finding disease risk loci, having successfully identified thousands of variants associated to complex phenotypes~\cite{manoliogwas,visscher12}. An important component of the GWAS analysis toolkit is genotype imputation, an approach that leverages publicly available data (e.g. 1000 Genomes~\cite{thousandgenomes}) to estimate genotypes at markers untyped in the study to increase power for finding new risk loci~\cite{imputationreview,Howie2012,beagle,Li10}. In addition to GWAS, genotype imputation is a key component of meta-analysis of studies that use different genotyping platforms, where SNPs that were genotyped in one study can be imputed in the other studies thus increasing the association power~\cite{estrada10,allen10,morris12,liu10}.

Many approaches for genotype imputation have been proposed, with methods based on Hidden Markov Models (HMM) showing the highest accuracy in simulations and empirical data\cite{imputationreview,Howie2012,beagle,Li10}.  However, privacy and logistic constraints often prohibit access to individual level genotype data thus precluding HMM-based imputation, whereas summary association statistics are becoming widely available.  For example, summary statistics are required to be publicly released for any GWAS published in Nature Genetics, and have been publicly released for many traits\cite{ngeditorial,shork2013summary}.   

In this work we propose methods for testing for association at SNPs untyped in the study when only summary association statistics are available at the typed SNPs. Unlike HMM-based imputation from individual-level genotypes, our proposed approach requires as input only the association statistics at typed variants. To accomplish this we approximate the distribution of association statistics at a given locus using a multivariate Gaussian. Previous studies have shown that a Gaussian approximation of linkage disequilibrium (LD) leads to accurate inference across a wide range of problems~\cite{multipop,Han2009,coneely2007,Wen2010,mcpeek2012blup}. In particular, ref. \cite{Wen2010}  highlighted the potential utility of Gaussian imputation methods for individual level and pooled data, but that study did not provide methods or software for imputation from summary association statistics (see Discussion).

Through extensive simulations based on 1000 Genomes data, we show that our approach is almost as powerful as the gold standard of HMM-based imputation from individual-level genotypes, and is able to avoid an increase in false-positive associations by accounting for the limited size of the reference panel. Our approach recovers 84\% (54\%) of the effective sample size for common ($>$5\%) and low-frequency (1-5\%) variants, versus 89\% (67\%) for HMM-based imputation, with a reduction in running time of several orders of magnitude. When summary information on the pairwise LD structure among typed variants in GWAS samples is made available (as is also recommended in other contexts\cite{lee2013general}), our method recovers 87\%(60\%) of the effective sample size, again with no increase in false-positive associations. 

We validate our approach using real GWAS data from WTCCC across 7 phenotypes as well as a height GWAS from the 1958 Birth Cohort (1958BC), where we show that Gaussian imputation from summary statistics recovers the same signal as HMM-based imputation from individual-level genotypes, with no increase in false-positive rate.  For example, we attain an average $\chi^2$ association statistic of 18.28 as compared to 19.17 for HMM-based imputation at the 227 published SNPs in the WTCCC data and 4.76 (vs. 4.55 for HMM-based imputation) at the 176 published SNPs in the 1958BC height data. 

For publicly available summary statistics from large meta-analyses of 4 lipid traits (triglycerides (TG), total cholesterol (TC), high density lipoprotein (HDL), low density lipoprotein (LDL)), we publicly release imputed summary statistics at 1000 Genomes SNPs, which could not have been obtained using previously published methods. We validate the accuracy of the imputed statistics across the 4 studies using a masking approach and show that we attain a correlation of 0.98 (0.95) to masked summary statistics for common (low-frequency) variants, consistent with  simulations.  {Finally, we explore the utility of imputed association statistics to functional enrichment analysis\cite{shork2013summary}. For the 4 lipid traits, we find that imputed data increases the magnitude and statistical evidence of enrichment at genic vs. non-genic loci, as compared to an analysis without 1000 Genomes imputation\cite{shork2013summary}.}

\section*{Methods}

\subsection*{Overview of Gaussian imputation of summary statistics}

We assume that summary association statistics consist of $z$-scores known to be normally distributed with mean 0 and variance 1 under the null model of no association. LD between SNPs $i$ and $j$ induces a covariance between their observed z-scores, according to the correlation $r_{ij}$ between the two SNPs. Thus, under null data the vector $Z$ of z-scores at all SNPs in a locus follows a Gaussian distribution, $Z  \sim  N(0,\Sigma)$, with $\Sigma$ being the correlation matrix among all pairs of SNPs induced by LD ($\Sigma_{ij} = r_{ij}$). In a given study we only observe z-scores at the typed SNPs ($Z_{t}$), with no information about untyped SNPs . We estimate $\Sigma$ using reference panels of haplotypes (e.g. 1000 Genomes) and analytically derive the posterior mean of $z$-scores at unobserved SNPs ($Z_{i}$) given $Z_{t}$ and $\Sigma$  (ImpG-Summary). We use the conditional variance to estimate the imputation accuracy ($r^2pred$), in a manner similar to the $r^2hat$ estimator in HMM-based imputation\cite{Li10}. 

The finite sample size of the reference panel adds statistical noise to the estimate of $\Sigma$. We account for noisy estimates at distant SNPs by employing a windowing strategy that models distant SNPs as uncorrelated. This strategy also leads to efficient computational runtime (a smaller matrix needs to be inverted for each window in the genome). In particular, we partition the genome into non-overlapping windows (e.g. 1Mb) and for each window independently, we estimate the LD matrix $\Sigma$ using the reference panel of haplotypes. To account for SNPs at the boundaries of these windows we include SNPs within a buffer around the window in the computation (e.g. 250Kb on either side). We also account for statistical noise in estimates at proximal SNPs by adding $\lambda I$ to the LD matrix estimated from the data $\Sigma$. This procedure is similar to ridge regression \cite{ridge} and can also be interpreted in a Bayesian context as adding a prior of $N(0,\lambda I)$ to the typed SNP coefficients\cite{trevor2001elements}. Accounting for this statistical noise is necessary to eliminate false-positive associations (see Results). 

The imputed $z$-scores (from the conditional Gaussian distribution) can be viewed as a linear combination of typed $z$-scores with weights pre-computed from the reference panel. Therefore the variance of the imputed $z$-score can be estimated on the basis of the weights and the LD among typed SNPs. We estimate the LD structure among typed SNPs using the reference panel as above (accounting for statistical noise in the reference panel) and normalize the imputed $z$-scores such that their theoretical variance under the null is 1 (in a real scan the observed variance may be greater than 1 due to polygenic effects\cite{yang2011genomic}). Since we use fixed window sizes the computation time of our procedure scales linearly with the total number of SNPs as the computations at each window can be performed in constant time (proportional to the square of the fixed number of SNPs in the window). The computation time can be further reduced by precomputing the inverse of the covariance matrix for each genotyping array platform at each window, although special requirement is required for typed SNPs that are removed by QC. In particular, matrix inversion should be repeated at windows where typed SNPs used in the imputation are removed by QC.

If summary LD statistics (pairwise LD among typed SNPs within a window in GWAS samples) are also available, they can be directly used to estimate the variance of each imputed $z$-score (which can be viewed as a linear combination of typed $z$-scores). This produces an accurate estimate of the expected variance under the null for the imputed $z$-scores with no need of adjustment for the statistical noise in the reference panel (ImpG-SummaryLD). This leads to well calibrated association statistics under the null with increased power relative to ImpG-Summary.

\subsection*{Association statistics in GWAS}

A standard test for association in GWAS is the normalized difference in frequencies between cases and controls (z-score $z$) defined as:  $$  z = \sqrt{N} \frac{f^+ - f^-}{\sqrt{2f(1-f)}} $$ where $f^+$ ($f^-$) denotes the frequency in cases (controls), $f$ is the overall frequency and $N$ the number of samples.  This statistic extends to continuous phenotypes by considering $\sqrt{N}$ times the correlation between the vector of  genotypes (0,1,2) and phenotype. In the case of imputed data, this statistic extends by using genotype dosages in the computation of the correlation of dosages to phenotype.  Linkage Disequilibrium (LD) between pairs of SNPs $s$ and $s'$ induces a correlation among the observed z-scores at these SNPs which can be expressed through the standard correlation coefficient $r(s,s')$.

\subsection*{Multivariate Gaussian approximation}

Similar to other works\cite{multipop,Han2009,coneely2007,Wen2010}, we approximate the full distribution of association statistics $Z$ at $n$ SNPs  in LD using a multivariate Gaussian distribution with probability density function depending on the mean $\mu$ and variance covariance $\Sigma$.  Let the vector $Z$ be partitioned into two components $Z_t$ and $Z_i$ corresponding with the typed and imputed SNPs, where $Z_t$ is a vector of size $m$ (assuming $m$ SNPs have been typed) and $Z_i$ has $n-m$ elements. Similarly, we will partition the mean vector and variance-covariance matrix into $(\mu_t,\mu_i)^T$ corresponding to the means at typed and imputed SNPs, covariances among imputed  ($\Sigma_{i,i}, [n-m\times n-m])$, covariances among typed and imputed  ($\Sigma_{i,t}, [n-m\times m])$) and  covariance among typed data  ($\Sigma_{t,t}, [m\times m])$). Then the conditional random variable $Z_i|Z_t$ follows an Gaussian distribution with mean $\mu_{Z_i|Z_t} = \mu_i + \Sigma_{i,t}\times {\Sigma}^{-1}_{t,t}\times (Z_t-\mu_t)$ and  covariance: $\Sigma_{i|t} = \Sigma_{i,i} -  \Sigma_{i,t}\times {\Sigma}^{-1}_{t,t}\times \Sigma^{'}_{i,t}$ .

\subsection*{Gaussian imputation of association statistics (ImpG-Summary)}

When modeling the variance covariance matrix $\Sigma$, we adopt a windowing strategy aimed at decreasing runtime (a smaller matrix needs to be inverted for each window in the genome) and at reducing statistical noise that can show distant SNPs to be correlated when the true is of no LD. In particular, we partition the genome into non-overlapping windows of 1Mb (with a buffer of 250Kb on either side to account for LD at boundaries). Let $\Sigma$ denote an estimate of the variance covariance matrix in the GWAS sample across both typed and imputed SNPs.   For each window independently, we estimate $\Sigma$ from the reference panel of haplotypes, with an adjustment for sampling noise (see below). Let $z_t$ be the set of observed z-scores restricted to current window. We impute $z_i$ as $z_{i|t} = \Sigma_{i,t}\times {\Sigma}^{-1}_{t,t}\times z_t$. To speed up computation, we can precompute $\Sigma^{-1}_{t,t}$ for all genotyping array platforms such that runtime is quadratic in the number of SNPs in  the window. For windows where QC has removed part of the typed SNPs used in imputation $\Sigma^{-1}_{t,t}$  needs to be re-estimated. Since the window length is fixed across the genome, the overall computational runtime can be thought of linear in the number of  SNPs (when ${\Sigma}^{-1}_{t,t}$ has been precomputed already).

The imputed z-scores at imputed SNP i $z_{i|t}$ can be viewed as a linear combination of typed z-scores $z_t$ with weights $W=\Sigma_{i,t}\times {\Sigma}^{-1}_{t,t}$ pre-computed from the reference panel. Let $A$ denote the variance covariance matrix among typed SNPs in the population.  Since $z_t$ follows a Gaussian distribution $N(0,A)$, it follows that $z_{i|t}$ has variance $W \times A \times W' $. Therefore,  we use  $\frac{z_{i|t}}{\sqrt{W \times A \times W'} }$ as the imputation z-score at imputed SNP i. To account for the statistical noise while also making sure that $\Sigma$ is invertible we adopt a procedure similar to ridge regression \cite{ridge} and use  $\Sigma = \Sigma^{unadj} + \lambda I$ in both $\Sigma_{i,t}$ and $\Sigma_{t,t}$ in the estimation of $W$ (we use $\lambda=0.1$ as default) (see  Tables S1,S2,S3 for results across other values of $\lambda$). We approximate $A$ with $\Sigma_{t,t}$ using LD information from reference panel (ImpG-Summary). 

An alternative is to use the true $A$, i.e. the summary LD statistics from the GWAS sample, if they are available; in this case, a more substantial adjustment for statistical noise in $\Sigma$ is not needed because $A$ is derived from the GWAS sample, and we set  $\lambda=0.001$ to make sure that $\Sigma$ is invertible in the estimation of $W$ (ImpG-SummaryLD). We do not use the summary LD statistics across typed SNPs in the sample for estimation of $W$ in ImpG-SummaryLD, to maintain consistency among pairwise LD statistics between typed and imputed SNPs. Software implementing the ImpG-Summary and ImpG-SummaryLD methods has been made publicly available (see Web Resources).

We propose a metric for imputation accuracy based on the variance of the conditional random variable $Z_i|Z_t$: we define $r^2pred = 1- \Sigma_{i|t}$.  Figure S1 shows that $r^2pred$  behaves very similarly to the standard imputation accuracy metric $r^2hat$\cite{Li10} (correlation of 0.90 to the true $r^2$ accuracy as compared to 0.92 for $r^2hat$).

\subsection*{Simulation framework}

We simulated data starting from the 381 diploid European individuals from the phase 2 release of the 1000 Genomes Project (June 2011)\cite{thousandgenomes}. The 381 individuals include 87 CEPH individuals of North European ancestry (CEU), 93 Finnish individuals from Finland (FIN), 89 British individuals from England and Scotland (GBR), 98 Tuscan individuals (TSI), and 14 individuals from the Iberian peninsula (IBS). Genotype calls and haplotypic phase had been previously  inferred from low-coverage sequencing (4x) using an imputation strategy that borrowed information across samples and loci\cite{thousandgenomes}. The set haplotypes were split at random between a set of 178 (number chosen to match the 89 samples of British ancestry) haplotypes used to build simulated data, and the other was used as an imputation reference panel. Starting from the simulation panel of haplotypes, we used hapgen\cite{hapgen2} to simulate 10,000 diploid individuals. All simulation results were generated over 50 distinct 1Mb regions  (total of 50Mb) randomly chosen across Chromosome 1 totaling 321,226 SNPs. For each of the SNPs with MAF greater than 1\% in the reference panel (133,025 in total),   we simulated case-control data sets by randomly choosing a subset of 1,000 controls, and then chosing 1,000 cases from the remaining samples so  that samples with 0:1:2 reference alleles have relative probabilities 1:$R$:$R^2$ of being chosen (for a given odds ratio $R$ ). For null simulations we randomly selected 1,000 samples as cases and 1,000 samples as controls.

\subsection*{WTCCC data set}

We examined data from the Wellcome Trust Case Control Consortium (WTCCC) phase I  comprising GWAS studies of  7 diseases:  Bipolar disorder (BD),  Coronary heart disease (CAD) ,  Crohn's disease (CD), Hypertension (HT), Rheumatoid arthritis (RA), Type 1 diabetes (T1D), Type 2 diabetes (T2D) (see  Table S4 for detailed sample sizes )\cite{wtccc}. We removed all SNPs that had  overall deviation from Hardy-Weinberg equilibrium at a p-value below 0.01. Then, we removed any SNP that had differential missigness (p-value $<$ 0.01) in any of the case-control cohort, overall missingness over 0.001, or minor allele frequency below 0.01. This yielded a total of 325,553 SNPs. We performed HMM-based imputation using  the pre-phasing approach of \cite{Howie2012}; we used HAPI-UR\cite{williams2012phasing} to infer haplotypes from genotypes and then ran IMPUTE2\cite{Howie2012} using default parameters on the inferred haplotypes (see Web Resources). Unless otherwise noted we filtered out imputed SNPs using an imputation accuracy cutoff of 0.6, as well as SNPs that had more than 5\% of the individual imputed calls missing at a posterior probability level of 0.9. This procedure yielded approximatively 4.7M SNPs  for the considered phenotypes (Table S4).

\subsection*{1958 Birth Cohort data}

The British 1958 birth cohort is an ongoing follow-up of all persons born in England, Scotland and Wales during one week in 1958. At the age of 44-45 years, the cohort were followed up with a biomedical examination and blood sampling\cite{strachan07}, from which a DNA collection was  established as a nationally representative reference panel (http://www.b58cgene.sgul.ac.uk/). Non-overlapping subsets of the DNA collection were genotyped by the Wellcome Trust Case-Control Consortium (WTCCC)\cite{wtccc}, the Type 1 Diabetes Genetics Consortium (T1DGC)\cite{barrett09} and the GABRIEL consortium\cite{moffatt10}. Genotyping by the WTCCC used both the Affymetrix 500K array and the Illumina 550K (version 1) array. Since the T1DGC used the Illumina 550K (version 3) array and GABRIEL used the Illumina 610K array, a combined dataset was created of SNPs in common across these three panels. SNPs were excluded from subsequent imputation if they had MAF$<$1\%, call-rate$<$95\%, HWE p-value$<$0.0001, or differences in allele frequency across the three deposits (p$<$0.0001 on pairwise comparisons). Pre-imputation phasing was performed using MACH\cite{Li10}. Imputations against the March 2012 release of 1000-genomes all-ethnicities reference haplotypes were performed using Minimac\cite{Howie2012}.  Associations of imputed allele dosages with standing height, as measured at the 44-45-year follow-up, were analysed using ProbAbel\cite{aulchenko10}.

\subsection*{Publicly available summary statistics for 4 lipid traits}

Publicly available GWAS summary data across four blood lipids phenotypes (triglycerides (TG), total cholesterol (TC), high density lipoprotein (HDL), low density lipoprotein (LDL)) was downloaded from public access websites~\cite{teslovich}. This data has been recently used in an study of overlap of GWAS findings and functional data\cite{shork2013summary}; all QC steps are described elsewhere~\cite{teslovich}. The data comprised roughly 2.7M summary statistics based on roughly 100,000 samples for each of the four phenotypes. To remove strand ambiguity we removed all A/T and C/G SNPs (roughly 15.4\% of all SNPs); we also removed all SNPs with meta-analysis sample sizes under 80,000, leaving approximately 2.0M SNPs for each of the phenotypes. We imputed to 1000Genomes using ImpG-Summary under three scenarios. In the first scenario, we removed 10\% of the SNPs at random. In the second scenario, we removed all SNPs not present on the Illumina 610 genotyping platform (approximately 600k in total). In both of these scenarios, we imputed from the remaining SNPs and assessed accuracy using the previously masked SNPs.  As a metric of accuracy we computed the correlation between imputed and previously masked association statistics. In the third scenario, we imputed from all 2.0M SNPs to obtain the summary statistics at 7.3M SNPs that we publicly release.  

\subsection*{Enrichment analysis for 4 lipid traits}

We used an analysis similar to \cite{shork2013summary} to quantify enrichment per classes of SNPs. We categorized each SNP according to its distance to genes using the all SNPs track (snp137)  from the UCSC genome browser (http://genome.ucsc.edu/cgi-bin/hgTables). All SNPs within an exon, up to 5Kb up-stream, up to 5Kb down-stream, located in the 3' UTR or in the 5' UTR were labeled as genic. SNPs with no annotation in the data were considered as being Intergenic. For each data set, we normalized the association statistics using genomic control attained only over the Intergenic SNPs \cite{shork2013summary}, followed by computation of average variance across SNPs within each functional class. We estimate the variance as the average of the squared association z-scores minus 1 \cite{shork2013summary}. We compared the magnitude of enrichment in association statistics across different functional classes within the same data set (either the public data or the imputed one) using the median of the Kolmogorov-Smirnov (KS) test statistic at 100 random draws each of 10,000 random SNPs across the genome.  This conservative computation avoids correlations due to LD and does not account for the larger number of SNPs in the 1000G imputed data, which would further increase statistical significance.

\subsection*{Gaussian imputation of individual genotypes}

Although we focus primarily on imputation of summary statistics, for completeness we also discuss Gaussian imputation when individual level data is available. We compare two different approaches.  The first approach is to apply ImpG-SummaryLD as described above, relying only on summary association statistics and summary LD statistics.  The second approach (which attains slightly worse results) is very similar to the approach proposed by \cite{Wen2010}.  As described in that study, we can impute allele frequencies and treat each genotype as a sample of size 2. Following \cite{Wen2010}, we set $\mu$ to be the observed allele frequency in the reference panel and $\Sigma[i,j]$ to be the covariance between SNP $i$ and $j$. Next we apply the same windowing approach above to each sample independently to impute individual level genotypes. Although rare, in practice Gaussian imputation can output values less than 0 or greater than 2;  we adjust these values to 0 and 2, respectively. As association statistic we use the $\chi^2$ 1 dof statistic $N\rho^2(G',\phi)$, where $N$ is the number of samples and $\rho^2(G',\phi)$ is the squared correlation between the vectors of imputed genotypes and the phenotype.

\section*{Results}

\subsection*{Simulations}

To explore the effectiveness of Gaussian imputation using summary statistics (ImpG-Summary and ImpG-SummaryLD), we simulated case-control data sets at various effect sizes across a wide range of SNPs (see Methods). We used the 762 European haplotypes from the 1000 Genomes Project (phase 1, June 2011 release), and restricted the analysis to 50 distinct 1Mb regions (total of 50 Mb, containing 133,025 SNPs with MAF $>$ 1\%) randomly chosen across Chromosome 1.  We randomly selected 178 haplotypes for the simulation panel data and the remaining haplotypes for the imputation reference panel. Starting with the 178 haplotypes of the simulation panel, we used hapgen\cite{hapgen2} to simulate GWAS data sets. For each of the SNPs with MAF greater than 1\% in the reference panel we simulated GWAS data over 1,000 cases and 1,000 controls at different effect sizes. To assess the performance of imputation at recovering the true association signal when present, we used the relative effective sample size, defined as the ratio of average imputed $\chi^2$ statistics at untyped SNPs vs. $\chi^2$ statistics computed from true genotypes.  Here $\chi^2$ statistics refer to the squared z-score, which has a $\chi^2$ with 1 degree of freedom distribution under the null hypothesis. We envision that real scans will restrict their analyses to variants with high estimated imputation accuracy ($r^2pred > 0.6$, analogous to the $r^2hat$ estimator in HMM-based imputation\cite{Li10},  Figure S1), but we computed the relative effective sample size with all values of $r^2pred$ included in order to provide an appropriate assessment of power.  However, we restricted most of our analyses of false-positives to accurately imputed variants ($r^2pred > 0.6$), as these are the variants that would be analyzed in a real scan.

We first explored the robustness of imputation from summary statistics. ImpG-Summary attains genomic control $\lambda_{GC}$ of $0.94$ (see Figure S2) with no increase in false positive rate at the tail of the distribution (see Tables S1,S2,S3). Although ImpG-Summary attains a slight deflation (due to the adjustment procedure that has the effect of shrinking the predictor weights), this is necessary to avoid false-positives. As expected from the conditional distribution, Gaussian imputation with no variance normalization is deflated ($\lambda_{GC}=0.86$) while the naive normalization that does not account for the statistical noise in the LD matrix is also susceptible to false positives (we observe a near 4-fold increase in p-values smaller than $10^{-4}$ as compared to a well-calibrated statistic in null data simulations (see  Tables S1, S2 and  Figure S2)). Recent work in parallel to ours has also investigated the use of Gaussian models for summary association imputation but do not propose an adjustment for the statistical noise in the reference panel\cite{lee2013dist}. We caution that adjustment for the statistical noise in the reference panel is required to avoid false positives when using our method (see Figure S2).  Likewise, our simulations indicate that the method of \cite{lee2013dist}, which does not adjust for statistical noise in the reference panel, is susceptible to false positives at these reference panel sizes (see Figure S3).  However, when pairwise correlations among typed SNPs are available from the GWAS data, the expected variance under null of the imputed statistics can be accurately estimated and used for normalization (ImpG-SummaryLD). This removes the need for adjusting the LD matrix estimated from the reference panel leading to distributed association statistics with no susceptibility to false-positives ($\lambda_{GC}=1.00$ ,  Figure S2). 

We next assessed the ability of ImpG-Summary to identify true positive associations by measuring the decrease in effective sample size. Table \ref{table:power} shows the relative effective sample size in 1000 Genomes simulations with target and reference haplotypes randomly sampled from 762 European haplotypes (i.e roughly matched for ancestry). As a gold standard for imputation accuracy, we used Beagle, an HMM-based method that requires individual-level data\cite{beagle,browning2009unified}.  Beagle has previously been shown to achieve similar accuracy as other HMM-based methods, with far superior accuracy compared to tagging-based imputation\cite{browning2009unified,marchini2010genotype}. At an odds ratio of 1.5, ImpG-Summary recovers 84\% (54\%) of the effective sample size for common ($>$5\%) and low-frequency (1-5\%) variants, versus 89\% (67\%) for Beagle imputation. Interestingly, when LD information among the typed variants from the GWAS is available, ImpG-SummaryLD recovers 87\% (60\%) of the effective sample size, nearly as high as Beagle.  Table \ref{table:power} also shows the decrease in effective sample size across a wide array of odds ratios showing that the results are robust to different effect sizes.  Thus, imputation from summary statistics can recover most of the association power available from GWAS with individual-level data.

We also tested the effect of a mismatch in ancestry between the reference haplotype panel and the target population. We simulated case-control GWAS using the GBR haplotypes for target samples and  the remaining 1000 Genomes European haplotypes as reference haplotypes. Table \ref{table:gbrpower} shows only marginal decreases in performance for each of HMM,  ImpG-Summary and ImpG-SummaryLD as compared to previous results, with no excess of false positives ( Table S2). 

We note that both ImpG-Summary and ImpG-SummaryLD are computationally very fast, with running times several orders of magnitude lower than HMM-based methods for imputation from individual-level genotypes. Table \ref{table:runtime} shows a reduction in running time of several orders of magnitude for ImpG-Summary as compared to HMM-based approaches. For example, for a GWAS with 10,000 samples, IMPUTE2 with pre-phasing\cite{Howie2012} takes $>$200 CPU days ($>$40 CPU days if pre-phasing time is not included), whereas ImpG-Summary takes less than one CPU day (this can be further sped up using weights precomputed from the 1000 Genomes reference panel), as does ImpG-SummaryLD.  The magnitude of the difference in running time will only increase with larger studies (such as the N=100,000 studies analyzed below), as the running time of ImpG-Summary is independent of the number of target samples while the running time of HMM-based imputation is linear in this quantity.  However, we note that all of the methods listed in Table \ref{table:runtime} can be parallelized across regions of the genome for faster wall-clock running time.

Although our work focuses on imputation from summary statistics, for completeness we also investigate Gaussian imputation when individual-level data is available. This has been proposed in \cite{Wen2010} and shown to achieve similar accuracy as HMM-based imputation in the context of HapMap 3 data. In simulations from 1000 Genomes we find that our implementation of the \cite{Wen2010} approach (see Methods) achieves slightly but significantly lower accuracy than ImpG-SummaryLD (see  Tables S7,S8) across a wide range of effect sizes. The slight improvement of ImpG-SummaryLD over Gaussian imputation using individual-level genotypes suggests that there is an advantage to phenotype-aware imputation. When individual-level genotypes are available, HMM-based imputation remains the approach of choice due to its slightly higher accuracy, but we recommend the use of ImpG-SummaryLD in preference to previous methods for performing Gaussian imputation when rapidly prioritizing regions for HMM-based analysis.

\subsection*{Application to WTCCC and height data sets}

We explored whether similar results could be attained in real empirical GWAS data. We validated our approach using a WTCCC study spanning 7 diseases\cite{wtccc} (roughly 2,000 cases for each disease and 3,000 shared controls genotyped on Affymetrix 500K array (see  Table S4)) and a study of height involving 6,500 individuals from the British 1958 birth cohort (1958BC) genotyped on the Affymetrix 6.0 array (see Methods). Starting from the real genotype data, we used as reference all 758 European haplotypes of the 1000 Genomes phase 2 data to accurately impute approximately $4.3$ million SNPs with minor allele frequency $\>1\%$ either using an HMM-based method (IMPUTE2 with pre-phasing\cite{Howie2012}) or ImpG-Summary (see  Table S4). 

We compared association statistics at accurately imputed SNPs with either the HMM-based method or using ImpG-Summary (with the latter assuming no access to individual level data). We observed an average correlation of 0.94 between the two set of association statistics for both WTCCC and 1958BC phenotypes (Figure \ref{fig:wtccc}), showing high similarity between the two approaches (see  Figure S4 for each WTCCC phenotype). {In general we observe that the QQ and Manhattan plots show  similar behavior for HMM-based association as compared to ImpG-Summary imputation emphasizing no excess of false positives when only summary data is used in imputation} (Figures  S5-S13). In some instances we observe differences that we hypothesize represent false positives for the HMM-based imputation, likely due to insufficient QC filtering for the HMM-based approach (see Figures S10-S12). Importantly, statistics at known associated SNPs from the NHGRI GWAS catalog for each of the considered phenotypes \cite{manoliogwas} (Table \ref{table:wtccc}) show similar association power across the two compared methods (e.g. an average $\chi^2$ of 19.17 for HMM-based imputation versus 18.28 across the WTCCC data and 4.55 versus 4.76 for the height phenotype) (Figure \ref{fig:known},  Table S5) .

\subsection*{Application to publicly available summary statistics for 4 lipid traits}

We investigated the performance of ImpG-Summary on publicly available summary association statistic data sets of 4 blood lipid traits\cite{teslovich}. This data has been imputed using HMM-based imputation to an average of 2.0M markers (see Methods). We imputed this data to 7.3M 1000 Genomes markers.  We randomly masked 10\% of the data, re-imputed using ImpG-Summary and assessed accuracy by comparing ImpG-Summary with the masked data. As expected we observe a high correlation between the two sets of summary statistics (correlation r=0.98 (0.95) at common (low-frequency) variants; see Figure \ref{fig:teslovich},  Figure S14). To quantify the expected accuracy when imputation is performed from array-based association statistics, we also masked all association statistics not present on a standard genotyping array and re-imputed using ImpG-Summary. We again observe a high correlation (r=0.97 (0.91) at common (low-frequency) variants; see Figure 3,  Figure S14) thus showing that our approach recovers association statistics similar to those obtained by HMM-based imputation requiring individual-level genotypes. We have publicly released imputed summary association statistics obtained using the full set of 2.0M markers, without masking (see Web Resources). As expected, we observed lower $\lambda_{GC}$ for ImpG-Summary data as compared to original data (e.g. 0.92 versus 0.98 for HDL phenotype, see  Table S9). 

\subsection*{Enrichment analysis for 4 lipid traits}

We categorized each SNP according to functional classes (see Methods). We performed genomic control correction using $\lambda_{GC}$ estimated from only the Intergenic SNPs, as in\cite{shork2013summary}. After normalization, we estimated the average excess variance for each functional class as the average square of the association z-score minus 1\cite{shork2013summary}. We observe that 1000 Genomes imputation using ImpG-Summary increases the average variance for each functional class, with Genic SNPs (and in some cases Intronic SNPs) showing larger increases than Intergenic SNPs (see Figure 4,  Figure S15) .  The increase in $\lambda_{mean}$ for each functional class, even after normalization by $\lambda_{GC}$, indicates that 1000 Genomes imputation increases the ratio $\lambda_{mean}$/$\lambda_{GC}$, i.e. causes true signals to be more concentrated at the tail of the distribution. 

1000 Genomes imputation using ImpG-Summary  increases statistical evidence of enrichment at Genic vs. Intergenic SNPs, both because the magnitude of the enrichment is larger and because of the increased number of SNPs.  We focus here on just the former effect by computing KS test statistics at random subsets of 10,000 SNPs, a conservative computation that avoids correlations due to LD (see Methods). 
  Across all 4 phenotypes, median KS test statistics were more significant in the 1000G imputed data vs. the original data set (e.g. 4.75E-08 versus 7.63E-05 for HDL; see  Table S10 for all phenotypes). This highlights the increased utility of the 1000G imputed summary statistics that we have publicly released for analyses of functional enrichment.

\section*{Discussion}

We have introduced an approach for imputation of association statistics at untyped variants directly from summary association statistics using publicly available reference panels of haplotypes such as 1000 Genomes \cite{thousandgenomes}, in contrast to widely used HMM-based approaches that require individual-level genotypes\cite{Howie2012}. Through extensive simulations and real data analyses we show that our approach is almost as powerful as imputation from individual-level genotypes (for both common and low-frequency variants) with no excess of false-positives. We have described a method that uses summary association statistics (ImpG-Summary), as well as a method that uses summary association statistics and summary LD statistics (ImpG-SummaryLD).  Because summary LD statistics are not currently widely shared, we expect that ImpG-Summary will be of greatest practical value in the immediate future.  However, the slightly higher power attained by ImpG-SummaryLD provides a motivation for sharing of summary LD statistics to become a widely accepted practice.  This is likely to also prove valuable in other settings, such as conditional analysis or rare variant testing\cite{yang2012conditional,lee2013general}.  

It is often the case that privacy and logistic constraints prohibit the sharing of individual-level data. On the other hand, summary association statistics from large scale association studies are often readily available\cite{shork2013summary,teslovich}, despite the fact that privacy concerns may extend to summary data\cite{sankararaman2009genomic,homer2008resolving}. For example, a recent study used publicly available summary association statistics over a wide range of phenotypes to draw inferences about the enrichment of disease-associated variants in several functional categories\cite{shork2013summary}. Using the methods introduced here, such analyses can be expanded to the set of all 1000 Genomes variants. In particular, we have publicly released imputed association statistics at 1000 Genomes variants for 4 lipid traits. We show that for these 4 lipid traits, 1000 Genomes imputed summary statistics show a consistently larger and more statistically significant signal of enrichment in genic vs. non-genic regions as compared to the original publicly available data.  Thus, 1000 Genomes imputed summary statistics can be used to increase power in studies of functional enrichment.

The Gaussian approximation for LD among SNPs has previously been employed in a wide range of problems~\cite{multipop,Han2009,coneely2007,Wen2010,mcpeek2012blup}.  We showed that an adjustment similar to ridge regression removed the false-positive associations in imputed summary statistics that occurred when unadjusted estimates of LD were used. As reference panels become larger, we expect a smaller adjustment factor to be needed thus increasing accuracy. Large reference panels of typed SNPs could potentially also be employed to reduce the adjustment factor needed for avoiding false-positives \cite{yang2012conditional}. Other recent works have proposed to reduce the computational burden of imputation using a technique similar to matrix completion, however that approach does not extend to imputation from summary statistics~\cite{chi2013genotype}.

The work of \cite{Wen2010} presented methods for Gaussian imputation from allele frequencies in cases and controls or from individual-level genotypes. There are many key differences between that  work and the current study.   First, we impute association statistics (i.e. $z$-scores) rather than allele frequencies.  For case-control traits, it is unclear how to use imputed allele frequencies in cases and controls \cite{Wen2010} to obtain association statistics that are robust to false-positives; for quantitative traits, imputation of allele frequencies does not apply. Thus, the methods and software of \cite{Wen2010} cannot be used to impute association statistics, as we have done here.  Second, we evaluate our approach in simulations based on 1000 Genomes data\cite{thousandgenomes}, assessing both power and  false-positive associations.  Third, we validate our approach using real empirical data across several GWAS involving both discrete and continuous phenotypes, including the 4 lipid traits for which we have publicly released imputed association statistics at 1000 Genomes variants.  We note that recent parallel work has also proposed to use summary statistics with reference panels of haplotypes for imputation (\cite{lee2013dist}; a related approach is proposed in \cite{hu2013meta}), but that work does not provide a strategy to address false-positive associations arising from the limited size of the reference panel, as we do here.

We conclude with several limitations for the approaches we presented here. First, when summary LD statistics from the study are not available, our adjustment procedure leads to a slight deflation of association statistics under null data. This could hamper efforts to assess  confounding due to population stratification or cryptic relatedness via genomic control\cite{genomiccontrol}.  However, it is now widely recognized that genomic control is not an effective approach for assessing confounding in large studies, due to the expected inflation from polygenic effects \cite{yang2011genomic}. Second, application of our approach to summary statistics with inappropriate levels of QC is a potential concern due to the possibility of introducing false positives. Therefore, we caution that appropriate QC should be performed on typed variants prior to estimation of summary association statistics, as is standard procedure in any GWAS.  Third, recent work has shown that low-coverage sequencing is a more powerful alternative to genotyping arrays per unit of cost invested~\cite{Pasaniuc2012,LiGR,flannick2012,NielsenNRG}. The extension of Gaussian imputation to low-coverage sequencing data remains a direction for future work.  Finally, as with all imputation approaches, the methods presented here are more accurate for common variants than for low-frequency variants.  Accuracy will be even lower at very rare variants, although GWAS involving single-variant associations are generally focused on common and low-frequency variants.

\section*{Supplemental Data}

Supplemental data includes  10 Tables and 15 Figures. 

\section*{Acknowledgments}

This research was supported by NIH grant R01 HG006399 (B.P., N.Z., N.P., A.L.P.), R03 CA162200 (B.P.) and R01 GM053275 (B.P.).  We are grateful to J. Yang and D. Reich for helpful discussions, and to P. de Bakker and the International HIV Controllers Study for sharing data from ref. \cite{Pasaniuc2012}. We acknowledge use of phenotype and genotype data from the British 1958 Birth Cohort DNA collection, funded by the Medical Research Council grant G0000934 and the Wellcome Trust grant 068545/Z/02. (http://www.b58cgene.sgul.ac.uk/). Genotyping for the B58C-WTCCC subset was funded by the Wellcome Trust grant 076113/B/04/Z. The B58C-T1DGC genotyping utilized resources provided by the Type 1 Diabetes Genetics Consortium, a collaborative clinical study sponsored by the National Institute of Diabetes and Digestive and Kidney Diseases (NIDDK), National Institute of Allergy and Infectious Diseases (NIAID), National Human Genome Research Institute (NHGRI), National Institute of Child Health and Human Development (NICHD), and Juvenile Diabetes Research Foundation International (JDRF) and supported by U01 DK062418. B58C-T1DGC GWAS data were deposited by the Diabetes and Inflammation Laboratory, Cambridge Institute for Medical Research (CIMR), University of Cambridge, which is funded by Juvenile Diabetes Research Foundation International, the Wellcome Trust and the National Institute for Health Research Cambridge Biomedical Research Centre; the CIMR is in receipt of a Wellcome Trust Strategic Award (079895). The B58C-GABRIEL genotyping was supported by a contract from the European Commission Framework Programme 6 (018996) and grants from the French Ministry of Research.

\subsection*{Web Resources}
ImpG-Summary, ImpG-SummaryLD, precomputed weights from the 1000 Genomes reference panel, and imputed summary association statistics at 1000G SNPs for 4 lipid traits: http://bogdanlab.pathology.ucla.edu \\
Beagle imputation:  http://faculty.washington.edu/browning/beagle/beagle.html\\
HAPI-UR: https://code.google.com/p/hapi-ur/\\

\bibliography{mybib}

\newpage
\section*{Figure Titles and Legends}

\begin{figure}[h]
\centering
\caption{HMM-imputed (x-axis) versus ImpG-Summary (y-axis) association statistics (z-scores) for the BD phenotype in WTCCC Data (left) and over height phenotype in 1958 Birth Cohort Data (right).  Results for all other WTCCC phenotypes can be found in  Figure S4.}
\label{fig:wtccc}
\end{figure}

\begin{figure}[h!]
\centering
\caption{HMM-imputed (x-axis) versus ImpG-Summary (y-axis) association statistics (z-scores) at known associated SNPs from NHGRI GWAS Catalog in WTCCC (left) and  height in 1958 Birth Cohort Data (right). }
\label{fig:known}
\end{figure}

\begin{figure}[h!]
\centering
\caption{HMM-imputed (x-axis) versus ImpG-Summary (y-axis) association statistics (z-scores) for the TG phenotype in the blood lipids data. Left denotes imputation of 10\% of the z-scores using the remaining 90\% while right shows imputation results starting from all variants present on the Illumina 610 array. Results for all blood lipids phenotypes can be found in  Figure S13. ImpG-Summary took ~4 CPU days for the 10\% data and under 10 CPU hours for the array-based imputation.}
\label{fig:teslovich}
\end{figure}

\begin{figure}[h!]
\centering
\caption{Average variance per SNP  (average association $z^2$ -1) binned by different functional classes for all 4 blood phenotypes. Top displays the
absolute numbers attained across the original data and the ImpG-Summary imputation to  1000 Genomes  (r2pred$>$0.8). Bottom figure shows the absolute difference between original data and 1000 Genomes imputed association statistics. }
\label{fig:enrichment}
\end{figure}


\newpage

\section*{Tables}
\begin{table}[h]
\begin{center}
\begin{tabular}{  c | c   c c c c  }
Method  & \multicolumn{5}{c}{Odds Ratio} \\
               &  1.0  & 1.2  & 1.5 & 1.7 & 2.0 \\
 \hline
   & \multicolumn{5}{c}{All SNPs} \\
Beagle& 0.999& 0.892& 0.872& 0.870& 0.868 \\
ImpG-Summary& 0.937& 0.835& 0.823& 0.827& 0.836 \\
ImpG-SummaryLD& 0.999& 0.872& 0.851& 0.852& 0.855 \\
 \hline
   & \multicolumn{5}{c}{Common SNPs ( over 5\%)} \\
Beagle& 0.999& 0.900& 0.885& 0.883& 0.881 \\
ImpG-Summary& 0.956& 0.850& 0.841& 0.845& 0.855 \\
ImpG-SummaryLD& 0.999& 0.882& 0.867& 0.868& 0.872 \\
 \hline
   & \multicolumn{4}{c}{Low frequency SNPs (1 to 5\%)} \\
Beagle& 0.997& 0.808& 0.667& 0.640& 0.620 \\
ImpG-Summary& 0.881& 0.685& 0.539& 0.512& 0.491 \\
ImpG-SummaryLD& 0.997& 0.768& 0.597& 0.565& 0.542 \\
\end{tabular}
\end{center}
\caption{Relative effective sample size at imputed SNPs (ratio of the average $\chi^2$ association statistics attained at 
imputed versus typed SNPs) in simulated case-control studies at different effect sizes (R$>$1). The column corresponding to R=1 shows the average $\chi^2$ association statistic under the null model of no association. }
\label{table:power}
\end{table}

\newpage
\begin{table}[h]
\begin{center}
\begin{tabular}{  c  |  c   c   }
Method  & Rand & Great Britain \\
\hline
  & \multicolumn{2}{c}{All SNPs} \\
Beagle& 0.872& 0.868 \\
ImpG-Summary& 0.823& 0.818 \\
ImpG-SummaryLD& 0.851& 0.845 \\
 \hline
   & \multicolumn{2}{c}{Common SNPs ( over 5\%)} \\
Beagle& 0.885& 0.880 \\
ImpG-Summary& 0.841& 0.835 \\
ImpG-SummaryLD& 0.867& 0.860 \\
 \hline
   & \multicolumn{2}{c}{Low frequency SNPs (1 to 5\%)} \\
Beagle& 0.667& 0.671 \\
ImpG-Summary& 0.539& 0.549 \\
ImpG-SummaryLD& 0.597& 0.603 \\
\end{tabular}
\end{center}
\caption{Relative effective sample size at imputed SNPs (ratio of the average $\chi^2$ association statistics attained at 
imputed versus typed SNPs) when imputation is performed in a random subsample of the 1000 Genomes
European data or  over only Great Britain haplotypes (odds ratio is set to 1.5). }
\label{table:gbrpower}
\end{table}

\newpage
\begin{table}[h]
\begin{center}
\begin{tabular}{  c |  c   c c }
Method & N=1,000 & N=10,000 & N=50,000 \\
\hline
IMPUTE1		& 893.8	& 8,937.5	& 44,687.5 \\
IMPUTE2 (sampling)		&100	&1,000	&5,000\\
IMPUTE2 (pre-phasing)		&4.2	& 41.7	&208.3\\
IMPUTE2 (pre-phasing)*		& 21.5	&215.3	&1,076.4\\
Beagle 			&250	&2,500	&12,500\\
ImpG-Summary	&0.4	&0.4	&0.4\\
ImpG-SummaryLD	&0.4	&0.4	&0.4\\
\end{tabular}
\end{center}
\caption{Estimated runtimes for 1000 Genomes imputation for various number of individuals (N) in imputation. 
Runtimes given in central processing unit (CPU) days needed to impute across the whole genome 
(11.6 million SNPs polymorphic in Europeans). Runtimes for all versions of IMPUTE extrapolated 
from Howie et al 2012\cite{Howie2012}. IMPUTE2 (pre-phasing)* includes GWAS phasing time of 
25 minutes per individual\cite{Howie2012}. Beagle runtime extrapolated from an average of 
3 CPU hours runtime for N=300 samples across a 5Mb window in the genome. ImpG-Summary takes under 10 hours for imputation starting
from 600k typed variants and under 4 CPU days for imputation from 2M typed variants with no pre-computation.
}
\label{table:runtime}
\end{table}

\newpage
\begin{table}[h]
\begin{center}
\begin{tabular}{  c |  c c   c c }
Phenotype & Number of SNPs & HMM $\chi^2$  & ImpG-Summary $\chi^2$ & Ratio \\
\hline 
Bipolar disorder(BD)				& 9	& 7.02	& 6.66	& 0.95\\
Coronary heart disease (CAD)		& 32	& 13.39	& 13.11	& 0.98\\
Crohn's disease (CD)			& 70	& 20.78	& 19.74	& 0.95\\
Hypertension(HT)				& 7	& 4.47	& 3.95	& 0.88\\
Rheumatoid arthritis (RA) 		& 22	& 20.36	& 19.00	& 0.93\\
Type 1 diabetes (T1D)			& 36	& 36.39	& 34.98	& 0.96\\
Type 2 diabetes (T2D)			& 51	& 12.08	& 11.45	& 0.95\\
\textbf{All WTCCC}				&\textbf{227}	&\textbf{19.17}	&\textbf{18.28}	&\textbf{0.95}\\
\\
\hline
\textbf{Height 1958 Birth Cohort}			&\textbf{176}	&\textbf{4.55}	&\textbf{4.76}	&\textbf{1.05}\\
\end{tabular}
\end{center}
\caption{Average association statistics ($\chi^2$) over known associated SNPs from NHGRI GWAS Catalog for the 8 studied phenotypes. 
The average across all SNPs except HLA region (chr6:20-35Mb) in WTCCC data consisting of 216 SNPs in total  of 16.02 for HMM versus 15.29 for ImpG-Summary.}
\label{table:wtccc}
\end{table}

\newpage
\begin{figure}[h]
\centering
\includegraphics[width=0.45\textwidth]{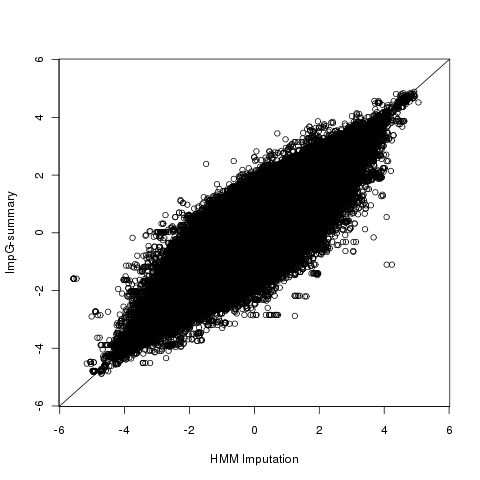}
\includegraphics[width=0.45\textwidth]{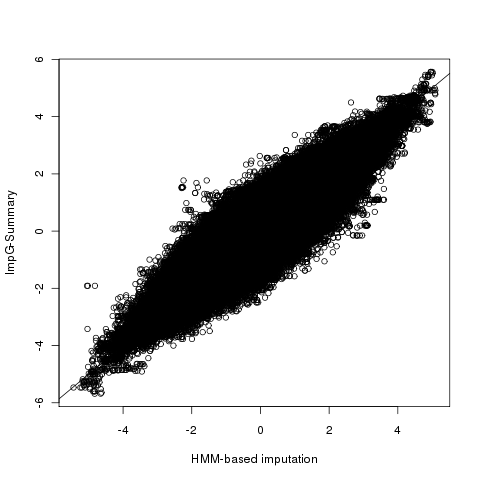}
\label{fig:1}
\end{figure}

\newpage
\begin{figure}[h]
\centering
\includegraphics[width=0.45\textwidth]{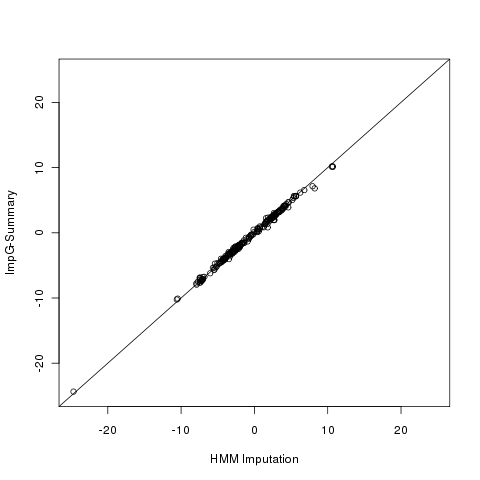}
\includegraphics[width=0.45\textwidth]{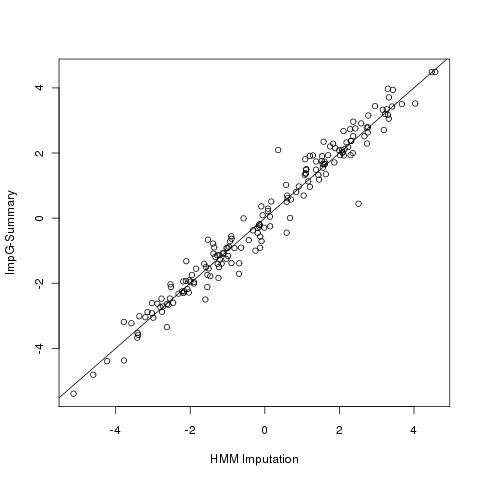}
\label{fig:2}
\end{figure}

\newpage
\begin{figure}[h]
\centering
\includegraphics[width=0.45\textwidth]{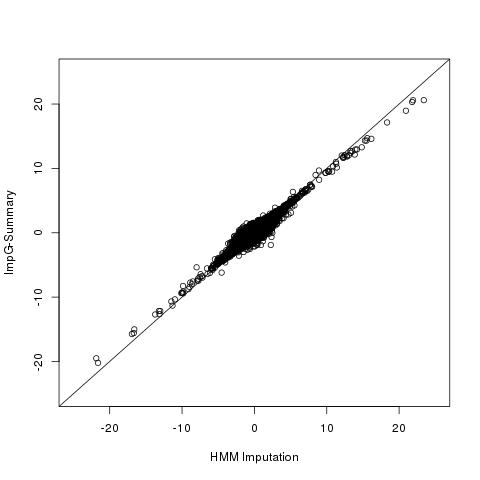}
\includegraphics[width=0.45\textwidth]{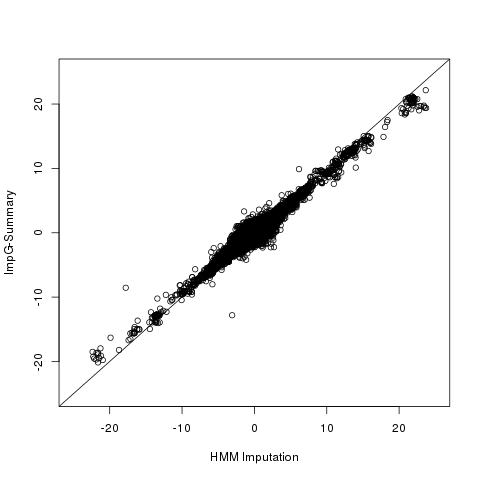}
\end{figure}

\newpage
\begin{figure}[h]
\centering
\includegraphics[width=0.6\textwidth]{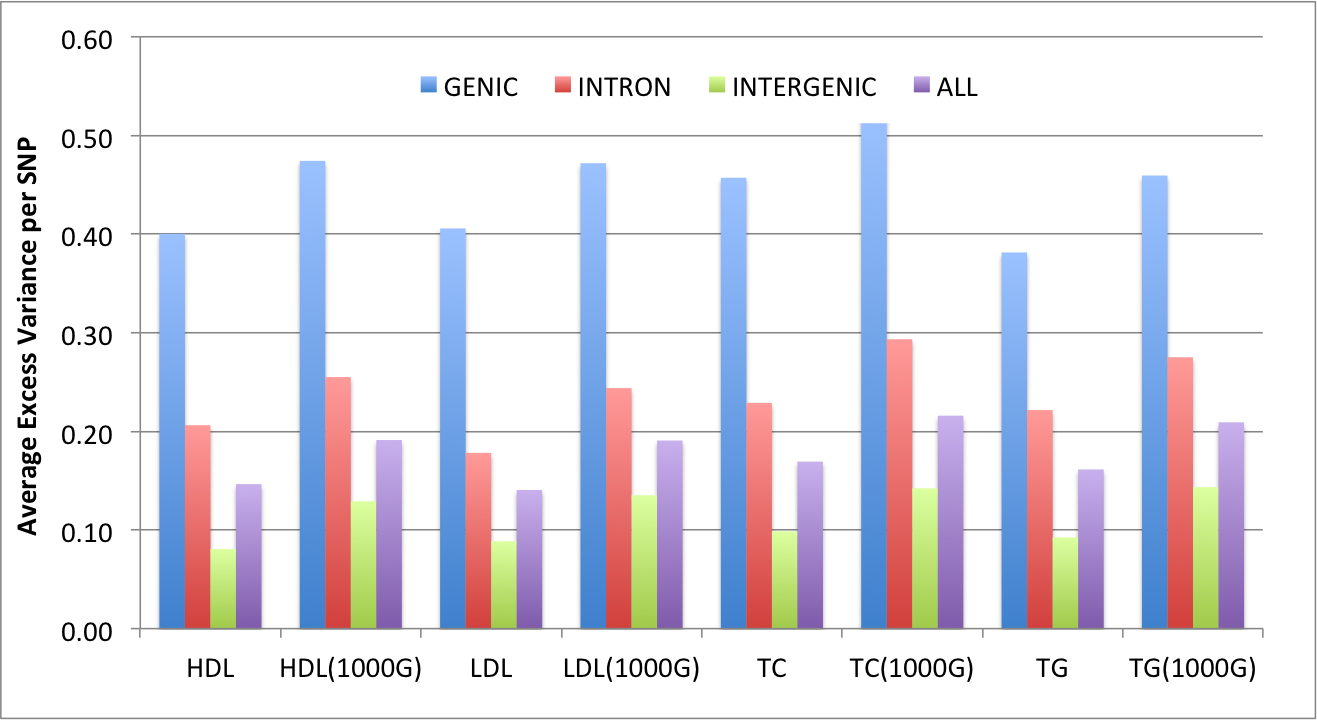}
\includegraphics[width=0.6\textwidth]{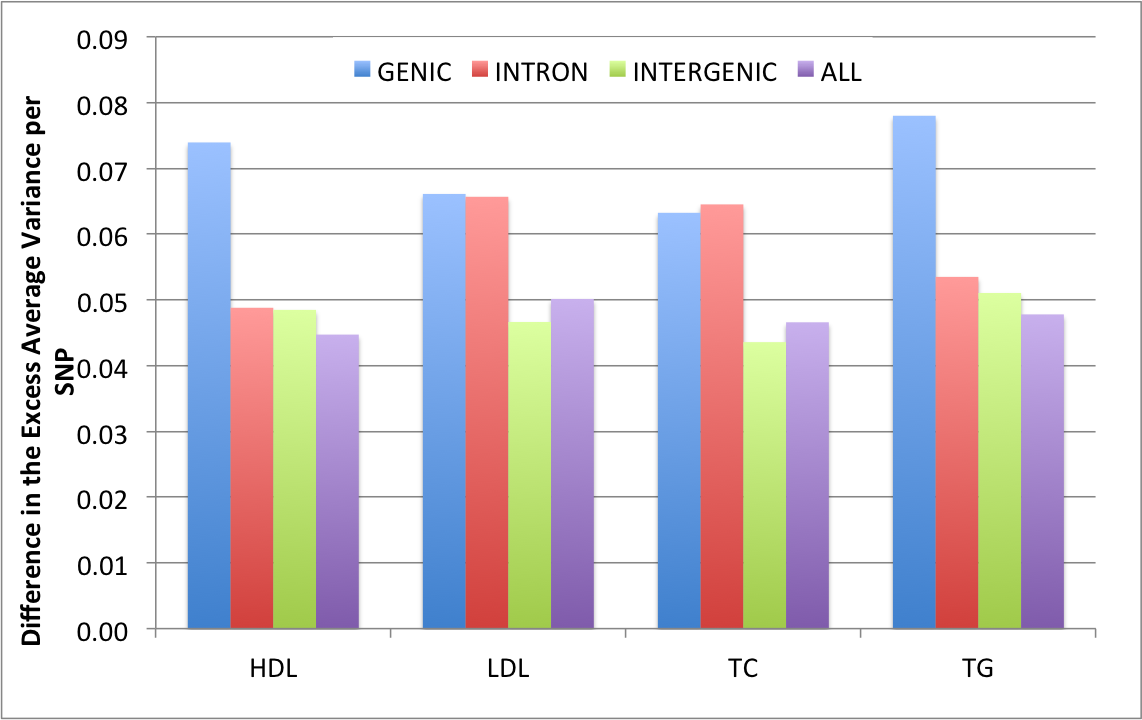}
\end{figure}

\end{document}